# Radiation Damping for Speeding-up NMR Applications

Gennady P. Berman[1], Michelle A. Espy[2], Vyacheslav N. Gorshkov[1,3],
Vladimir I. Tsifrinovich[4], and Petr L. Volegov[2]

[1] T-4, Theoretical Division, MSB213, LANL, Los Alamos, NM 87544
[2] P-21, Applied Modern Physics, D454, LANL, Los Alamos, NM 87544
[3] National Technical University of Ukraine "KPI," 37 Peremogy Av., Bldg 7, Kiev-56, 03056 Ukraine
[4] Department of Applied Physics, Polytechnic Institute of NYU, 6 MetroTech Center, Brooklyn, NY, 11201

**Abstract**

We demonstrate theoretically and numerically how to control the NMR relaxation rate after application of the standard spin echo technique. Using radiation damping, we return the nuclear magnetization to its equilibrium state during a time interval that is negligible compared to the relaxation time. We obtain an estimate for optimal radiation damping which is consistent with our numerical simulations.

## 1. Introduction

The problem of low repetition rate is one of the most important in many NMR applications [1]. There are two main approaches for controlling the nuclear magnetic relaxation [1-6]. The first is the application of an additional *rf* pulse in the process of the echo formation. This pulse can return the nuclear magnetization to its equilibrium position. However, in order to implement this method, the phase of the *rf* phase must be accurately adjusted relative to the phase of the transverse magnetization. The second method relies on the radiation field created by the resonant LCR coil interacting with the nuclear magnetization. In this method, the phase of the *rf* field produced by the coil is automatically adjusted relative to the transverse magnetization. However, the effect of the radiation field is, normally, small because the radiation field is much smaller than the *rf* field.

In this paper we demonstrate that, by increasing the duration of the spin echo and optimizing the coil inductance, one can restore the equilibrium position of the nuclear magnetization during the time of the echo formation, which is negligible compared with the relaxation time. Our consideration was done in the "dynamical" regime, when the characteristic time of the relaxation in the coil that produces the relaxation damping is much larger than the spin echo time. Our approach is valid for both, weak and strong magnetic fields. We obtained an analytical estimate for the optimal conditions for radiation damping, consistent with our numerical simulations.

## 2. Fast radiation damping with low-field setup

Consider a sample containing an ensemble of protons. This sample can interact with four coils. (See Fig. 1.) Coil "0" produces a permanent non-uniform magnetic field, $B_0$, in the positive *z*-direction. Coil "$1_x$" produces an oscillating field $B_{1x}$ (*rf*-pulses) along the *x*-axis. It is also used for NMR detection (with weak radiation damping). Coil "$1_y$" produces an oscillating field $B_{1y}$ (*rf*-pulses) along the *y*-axis. It also can be used for NMR detection (with weak radiation damping). Coil 2 produces the oscillating field, $B_2$, along the *x*-axis, which causes radiation damping. Note that we use two coils, $1_x$ and $1_y$, in order to produce circularly polarized *rf*-pulses. This is especially important if one is going to use an ultra-low permanent magnetic field, $B_0$. In the opposite case of high permanent magnetic field, instead of the circular polarized *rf*-pulses one can safely apply linearly polarized *rf*-pulses. In this case one of the coils ($1_x$ or $1_y$) can be removed.

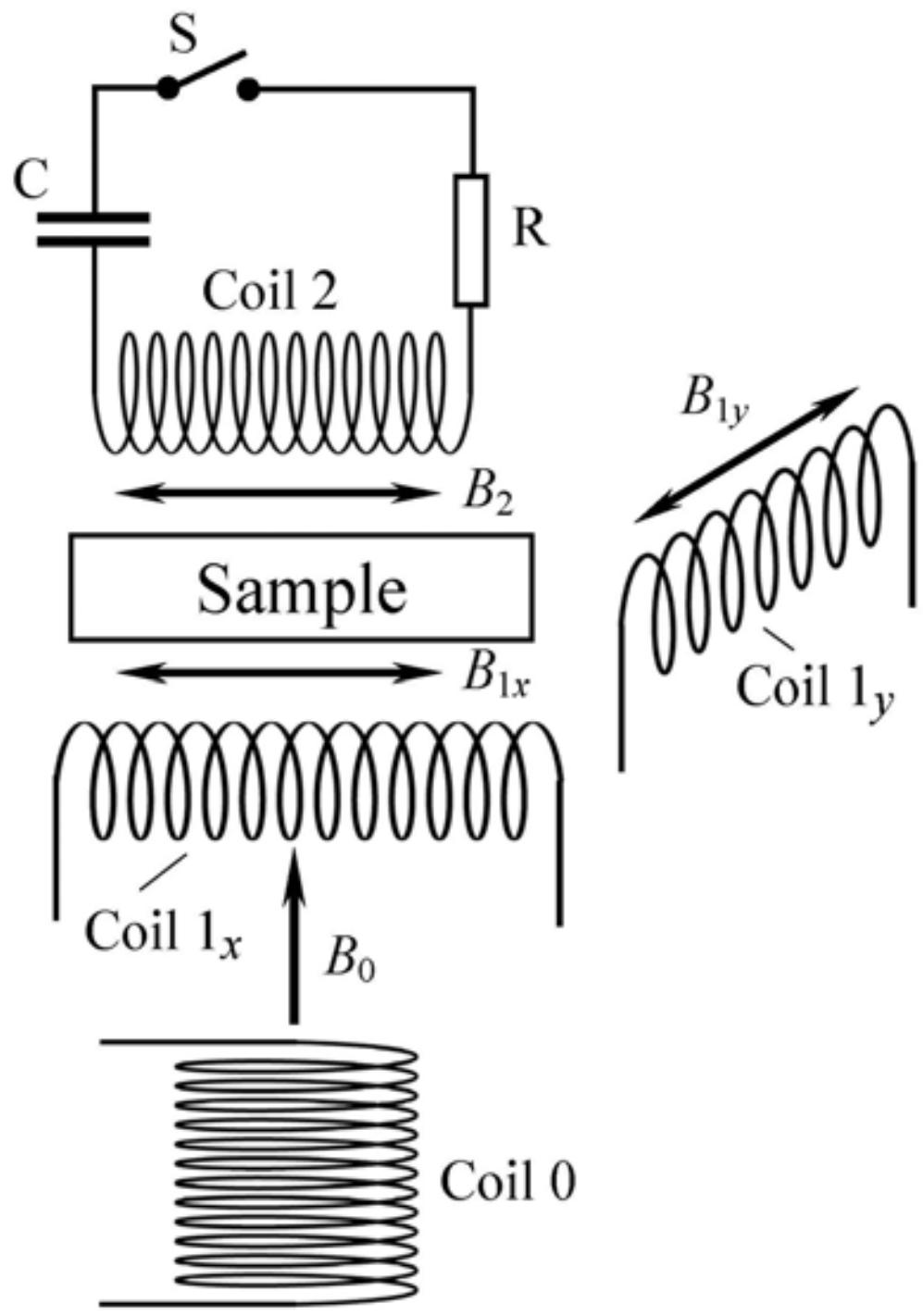

Fig. 1. Suggested setup.

We propose the following scheme:

1. At t = 0 a rectangular *rf* π/2-pulse of duration, say, 10 μs is applied to the sample. At time t = 1010 μs, a rectangular *rf* π-pulse of duration 10 μs is applied to the sample. At time t = 3020 μs, a rectangular *rf* π-pulse of duration 10 μs is again applied to the sample.

2. After the first pulse one observes a decaying signal of magnetic induction. Between the second and the third pulses one observes a spin echo. After the third pulse one again observes a spin echo. (See Fig. 2.)

3. Coil 3 is a part of an LCRS-circuit (S stands for a switch) with capacitance, C, and resistance, R. The switch, S, is closed (turned on) at t = 3030 μs, i.e. after the third pulse. Before the third pulse, coil 3 does not influence the spin system. During the duration of the second spin echo, the radiation damping due to coil 3 drives the nuclear magnetization toward its equilibrium state.

Below we write the equations of motion for $0 < t < 10$ μs together with definitions of notations:

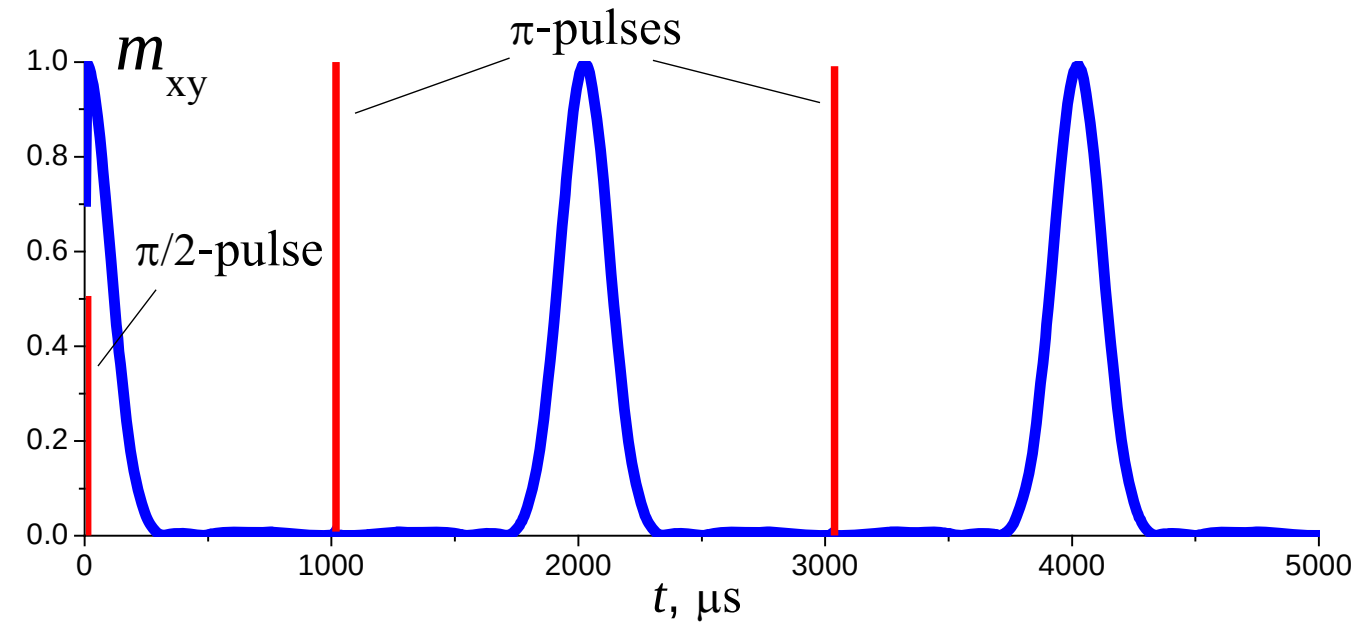

Fig. 2. Suggested spin echo sequence. $\Gamma = 10^4 s^{-1}$.

$$\begin{aligned}
&\dot{m}_x = m_y b_0 - m_z b_{1y}(t) - m_x / T_2, \\
&\dot{m}_y = m_z b_{1x}(t) - m_x b_0 - m_y / T_2, \\
&\dot{m}_z = m_x b_{1y}(t) - m_y b_{1x}(t) + \left(1 - m_z\right) / T_1, \\
&\left\langle \vec{m}(t) \right\rangle = \int_{-\infty}^{\infty} db_0 \vec{m}(b_0, t) g(b_0), \\
&b_{1x}(t) = a_1 \cos\left(\left\langle b_0 \right\rangle t\right), \\
&b_{1y}(t) = -a_1 \sin\left(\left\langle b_0 \right\rangle t\right), \\
&m_x(0) = m_y(0) = 0, \\
&m_z(0) = 1, \\
&b_0 = \gamma B_0, b_{1x} = \gamma B_{1x}, b_{1y} = \gamma B_{1y} \\
&g(b_0) = \left(1 / \sqrt{2\pi}\Gamma\right) \exp\left\{-\left(b_0 - \left\langle b_0 \right\rangle\right)^2 / 2\Gamma^2\right\}.
\end{aligned} \tag{1}$$

Here and below the equations of motion are written for a normalized dimensionless magnetic moment; $b_0$, $b_{1x}$, and $b_{1y}$ denote the corresponding magnetic fields, $B_0$, $B_{1x}$, and $B_{1y}$ in frequency units of $s^{-1}$; $g(b_0)$ is the Gaussian distribution for a permanent magnetic field, $b_0$, with the average value $\left\langle b_0 \right\rangle$; $\gamma$ is, for definiteness, the proton gyromagnetic ratio. In our numerical experiments described below, we used an averaging in Eqs (1) over an ensemble that included 25 thousand elementary magnetic moments, $\vec{m}^{(i)}$.

In our simulations, we use the following parameters: $\Gamma = 3.5 \times 10^3 s^{-1}$ and $\Gamma = 10^4 s^{-1}$, $B_0 = 1mT$, $T_1 = T_2 = 1$ s, the amplitude, $a_1$, of a π/2-pulse is found from the condition:

$$a_1 \cdot (10 \mu s) = \pi / 2. \tag{2}$$

$\langle B_0 \rangle$ = 1 mT.

The equations of motion for 10 < t < 1010 μs are:

$$\begin{aligned} \dot{m}_x &= m_y b_0 - m_x / T_2, \\ \dot{m}_y &= -m_x b_0 - m_y / T_2, \\ \dot{m}_z &= -(1 - m_z) / T_1. \end{aligned} \tag{3}$$

The equations of motion for 1010 < t < 1020 μs are:

$$\begin{aligned} \dot{m}_x &= m_y b_0 - m_z b_{1y}(t) - m_x / T_2, \\ \dot{m}_y &= m_z b_{1x}(t) - m_x b_0 - m_y / T_2, \\ \dot{m}_z &= m_x b_{1y}(t) - m_y b_{1x}(t) + (1 - m_z) / T_1. \end{aligned} \tag{4}$$

Here the amplitude, $a_1$, of the π-pulse is found from the condition:

$$a_1 \cdot (10 \mu s) = \pi. \tag{5}$$

The equations of motion for 1020 < t < 3020 μs are equations (3).

The equations of motion for 3020 < t < 3030 μs are equations (4).

The equations of motion for 3030 μs < t are the following:

$$\begin{aligned} \dot{m}_x &= m_y b_0 - m_x / T_2, \\ \dot{m}_y &= m_z b_2(t) - m_x b_0 - m_y / T_2, \\ \dot{m}_z &= -m_y b_2(t) + (1 - m_z) / T_1, \\ b_2(t) &= \gamma \mu_0 n I, \\ I / C &+ R\dot{I} + L\ddot{I} + \mu_0 \eta n V_c M \langle \ddot{m}_x \rangle = 0. \end{aligned} \tag{6}$$

Additional initial conditions at t = 3030 μs for a current in the coil 2 are:

$$I = \dot{I} = 0. \tag{7}$$

Here (for coil 2) $n$ is the number of turns per unit length, and $V_c$ is the volume of the coil; η is the filling factor (the ratio of the sample volume to the coil volume); and $M$ is the nuclear magnetization of the sample.

Our objective is to find the optimal conditions for a rapid return of the nuclear magnetization to the state, $m_z = 1$. We varied the parameter of the Gaussian distribution Γ, the resistance $R$, and the inductance, $L = \mu_0 n^2 V_c$, keeping the resonant condition $(LC)^{-1/2} = \langle b_0 \rangle$. Also, we use the following values of the parameters:

$$\alpha = \mu_0 \eta n V_c M = 10^{-6} \text{ Wb}, \qquad \beta = \gamma \mu_0 n = 3.36\text{x}10^6 \text{ rad/C}. \tag{8}$$

These parameters are computed using the values

$$\begin{aligned} &\gamma / 2\pi = 42.6 MHz / T, \\ &\mu_0 = 4\pi \times 10^{-7} H / m, \\ &n = 10^4 m^{-1}, \\ &\alpha = \mu_0 \eta n V_c M, \\ &\mu_0 M = 1.33 \times 10^{-7} T, \\ &V_c = 750 cm^3. \end{aligned} \tag{9}$$

## 3. Results of numerical simulations

The sequence of echo signals described above was simulated numerically. The dynamics of the average transverse magnetic moment, $m_{xy}(t) = \left( \langle m_x(t) \rangle^2 + \langle m_y(t) \rangle^2 \right)^{1/2}$, with no LCRS circuit is presented in Fig. 3.

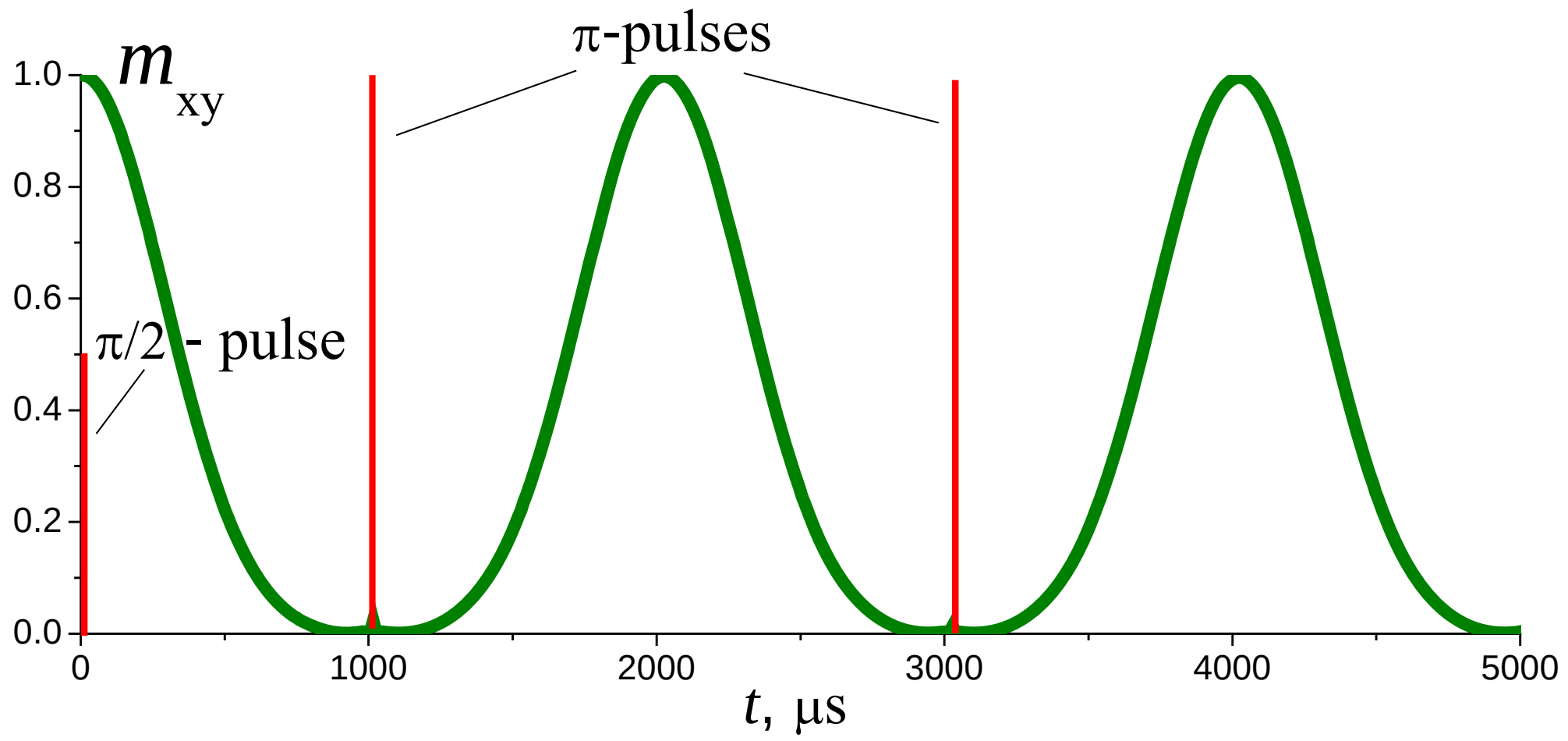

Fig. 3. Dynamics of $m_{xy}(t)$ for $\Gamma = 3.5\text{x}10^3$ s$^{-1}$.

Then, the numerical experiment for time, $t > t_3 = 3030 \mu s$, was done taking into account the LCRS circuit assuming that the switch is closed.

The physical process of fast radiation damping (FRD) can be described as follows. The induced transverse magnetic field, $B_{x2}(t)$, oscillating with the frequency, $\langle b_0 \rangle$, initially drives the nuclear magnetization toward the *z*-axis. The variation of the envelope, $m_{xy}(t)$, at the initial stage of this process is caused by two factors:

(i) the spin echo effect which increases the value of $m_{xy}(t)$;

(ii) the rotation of the magnetization vector in the field, $B_{x2}(t)$, which decreases the value of $m_{xy}(t)$.

The second factor is defined at the final stage of the FRD process.

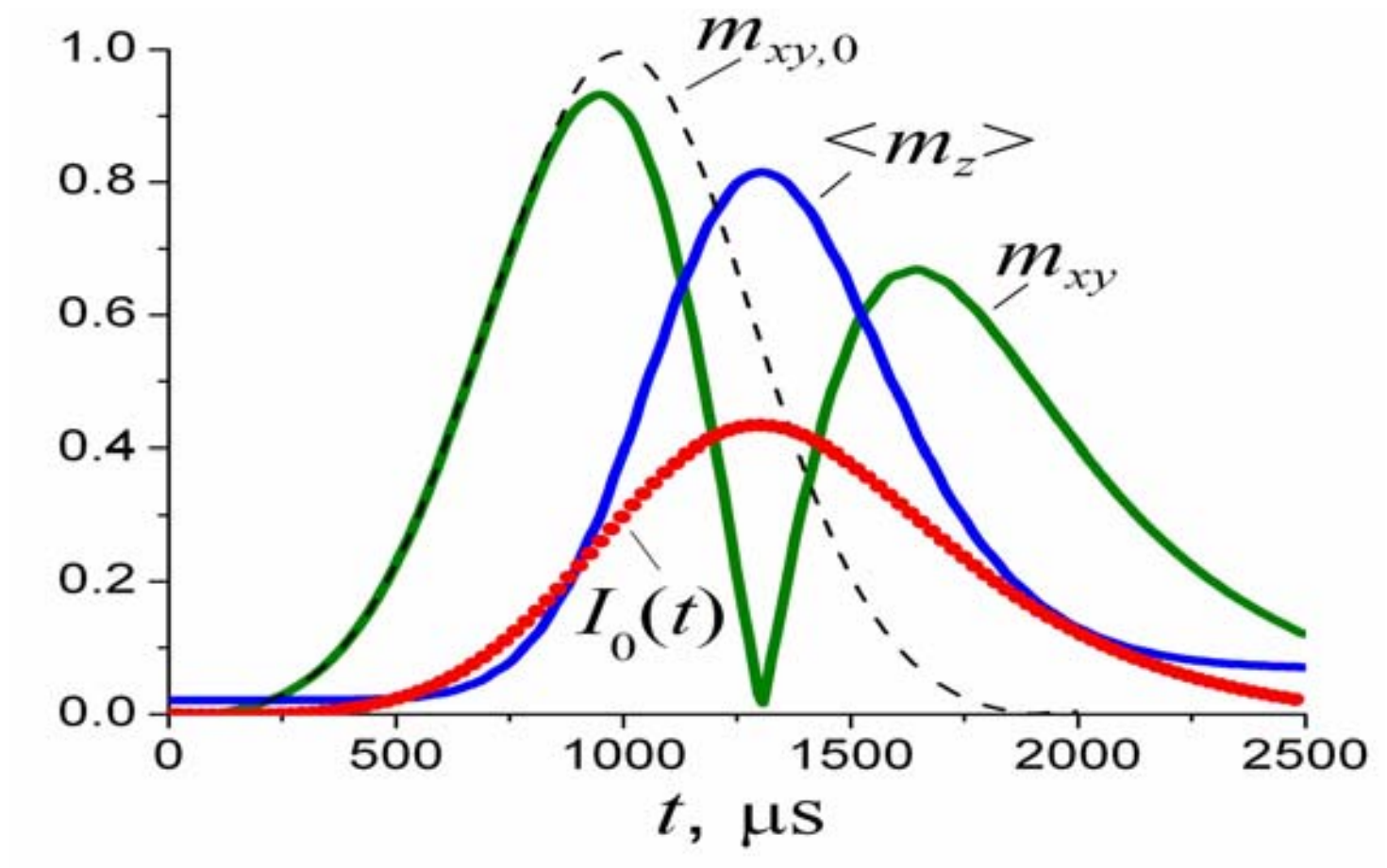

Fig. 4. The dynamics of the system after the second $\pi$-pulse for Γ = 3.5x10$^3$ s$^{-1}$, $Q = 3000$ and $L = 0.15H$. The time is counted here and below from $t_3 = 3030\mu s$. The amplitude of current, $I_0(t)$, is presented in *mA*. $m_{xy,0}(t)$ is the transversal magnetization when switch in LCRS circuit is open.

One could assume that at $\langle m_z(t) \rangle \sim 1$ , $\langle m_x(t) \rangle \to 0$ the state of the system would approach the equilibrium as $B_{x2}(t) \to 0$. In fact, the maximal values of $\langle m_z(t) \rangle$ (and, correspondingly, the smallest values of $m_{xy}(t)$) correspond to the maximal value of the current in the coil 2 (see Fig. 4). As a result, the vector of magnetization continues to rotate in the field, $B_{x2}(t)$, and the value of $\langle m_z(t) \rangle$ decreases. At first, this leads to an increase of the transverse magnetization, $m_{xy}(t)$, which then disappears due to the inhomogeneity of the field, $b_0(\vec{r})$. (See Fig. 4.)

The inductance of the LCRS circuit chosen in Fig. 4 is not optimal. Our numerical simulations show that the optimal condition for the FRD can be described by the following formula:

$$L_{opt} = \beta \frac{\langle b_0 \rangle \alpha}{4\pi} \tau_{echo}^2 . \qquad (10)$$

This relation can be obtained from simple considerations. (See Appendix.) For the field, $B_0 = 1mT$, $\alpha = 10^{-6} Wb$, $\beta = 3.36 \times 10^6 rad/C$, and $\tau_{echo} \approx 1ms$ (see Fig. 3) the relation (10) leads to $L_{opt} \approx 0.072H$. The corresponding optimal capacitance can be found from the resonance condition $LC = \langle b_0 \rangle^{-2}$. Substituting, $L = \mu_0 n^2 V_c$, we can rewrite the condition (10) for optimal radiation damping in the form:

$$\gamma \eta \mu_0 M \langle b_0 \rangle \tau_{echo}^2 / 4\pi = 1 . \qquad (11)$$

The values of our parameters (9) are chosen to match this condition. In applications, for a given values of $\gamma$ and $\eta$, one can vary $M, \langle b_0 \rangle$, and $\tau_{echo}$ by changing the pre-polarization field, the mean field $\langle B_0 \rangle$, and the non-uniformity of $B_0$, correspondingly.

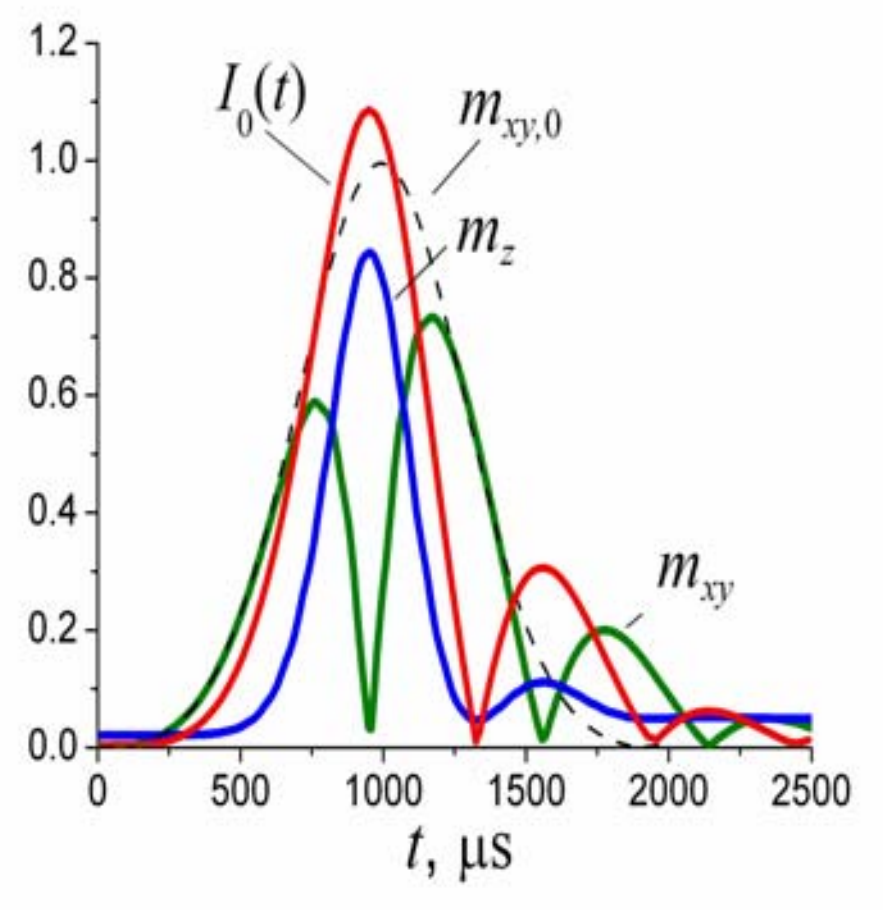

Fig. 5

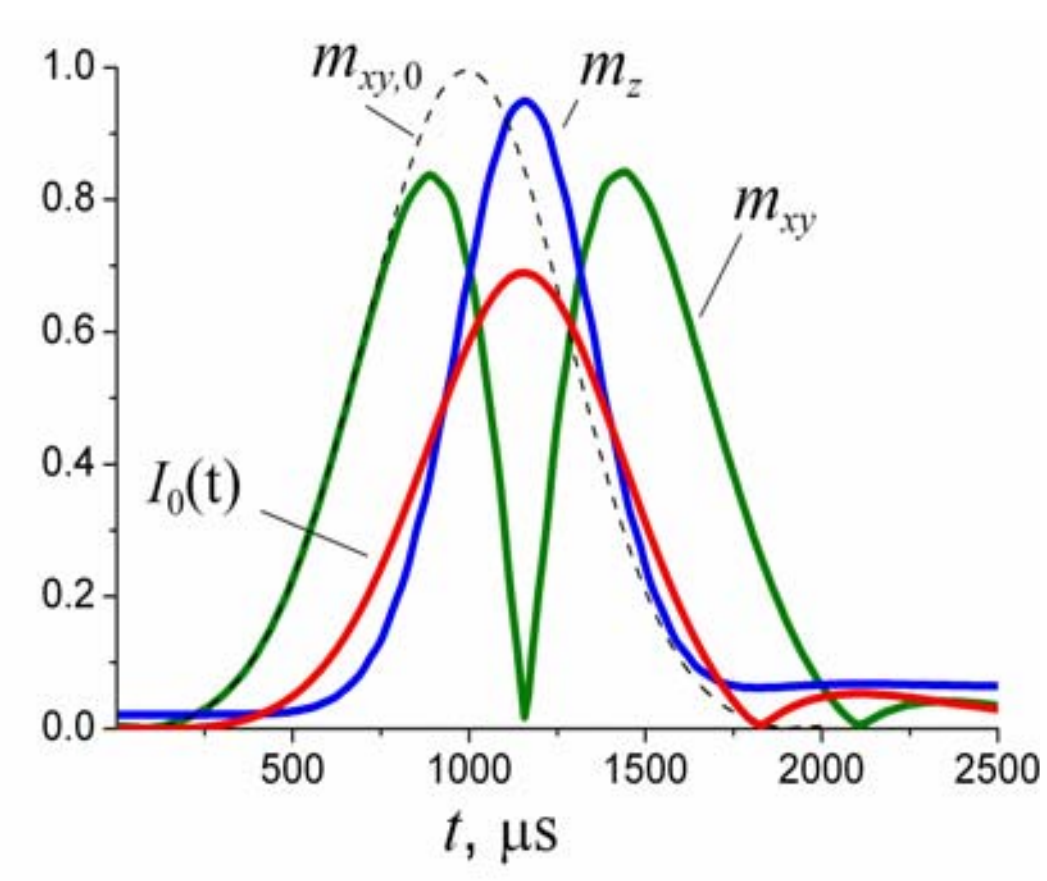

Fig. 6

The dynamics of the system after the second $\pi$-pulse for $Q = 3000$ and $\Gamma = 3.5x10^3 s^{-1}$. The amplitude of the current, $I_0(t)$ in $mA$. $L = 0.025H$ (Fig. 5) and $L = 0.075H$ (Fig. 6).

Our estimate (10) appears to be very good, as the results of numerical experiment demonstrate. In Fig. 5, the results of the numerical simulations are presented when the current in the circuit increases with advancing rate relative to the optimal regime. The maximal effect of FRD is realized at $L \approx 0.075H$ (see Fig. 6). In this case, the increase of $m_{xy}(t)$ in the process of spin echo formation is maximally used. As the results of numerical simulations demonstrate, the optimal choice of $L$ leads to the symmetric dependence of $m_{xy}(t)$ relative to the moment of time when $I_0(t)$ is maximal. Note that the FRD is not very sensitive to the value of $L$. As one can see from Figs. 4-6, changing of the inductance by a factor 6 does not significantly affect the FRD.

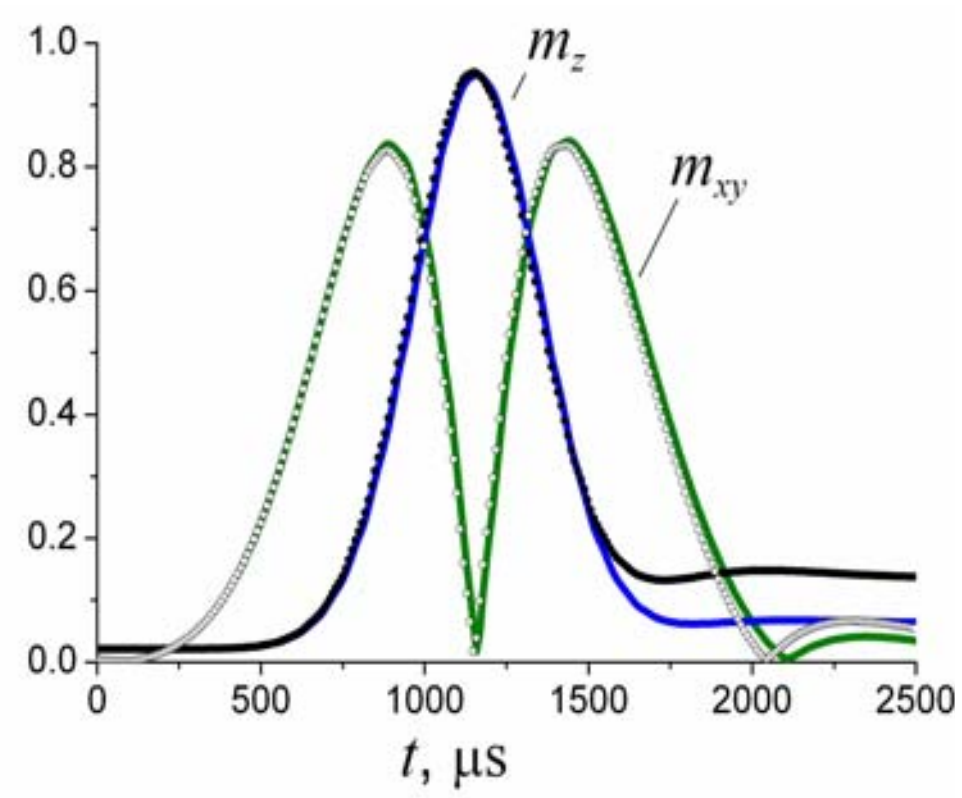

Fig. 7. The dynamics of the system for a given parameter, $\alpha$, and the inductance, $L$. Continuous curves: $Q = 3000$, dotted lines: $Q = 1000$.

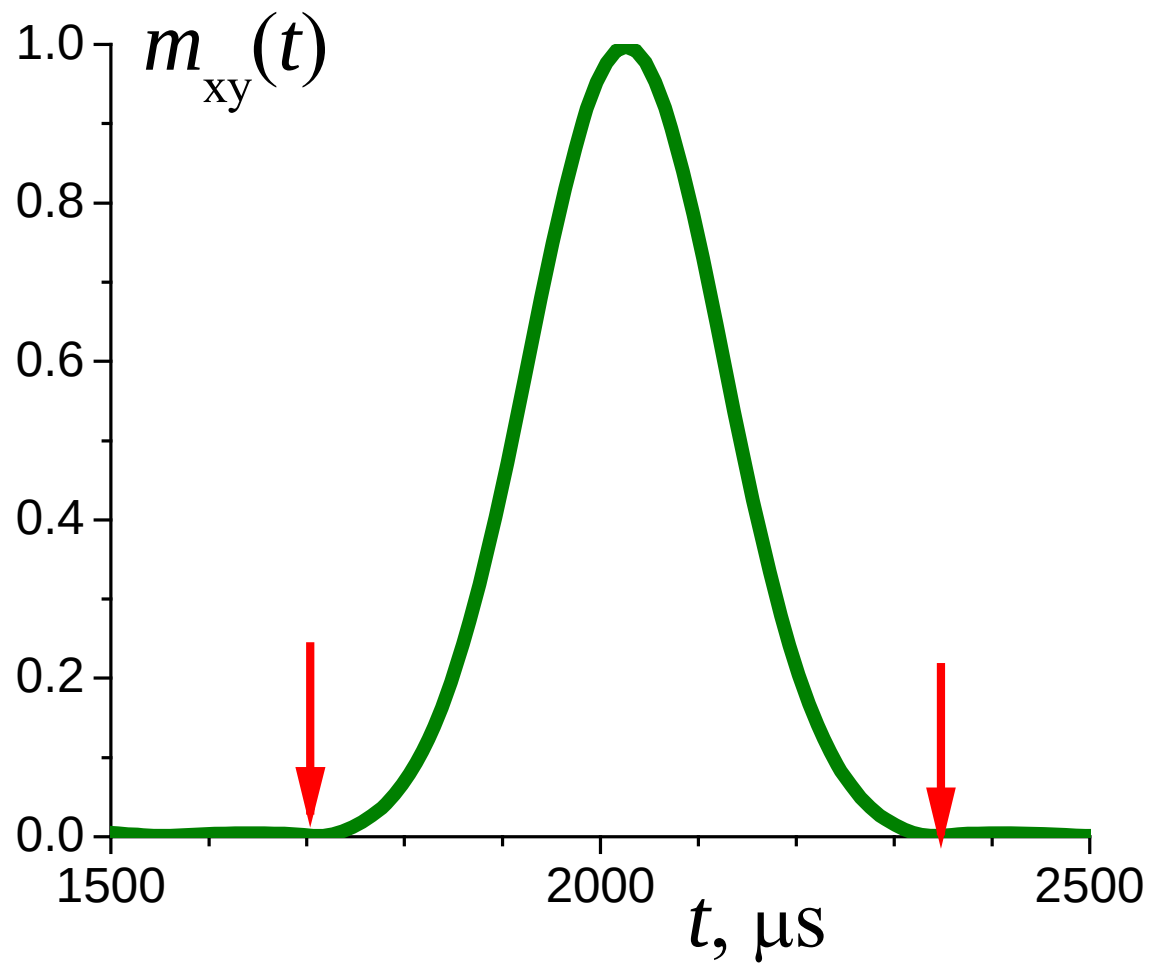

Fig. 8. Spin echo of the reduced duration. (This is a fragment of Fig. 2.) $\tau_{echo} \approx 0.325ms$ is half of the time-interval between the red arrows.

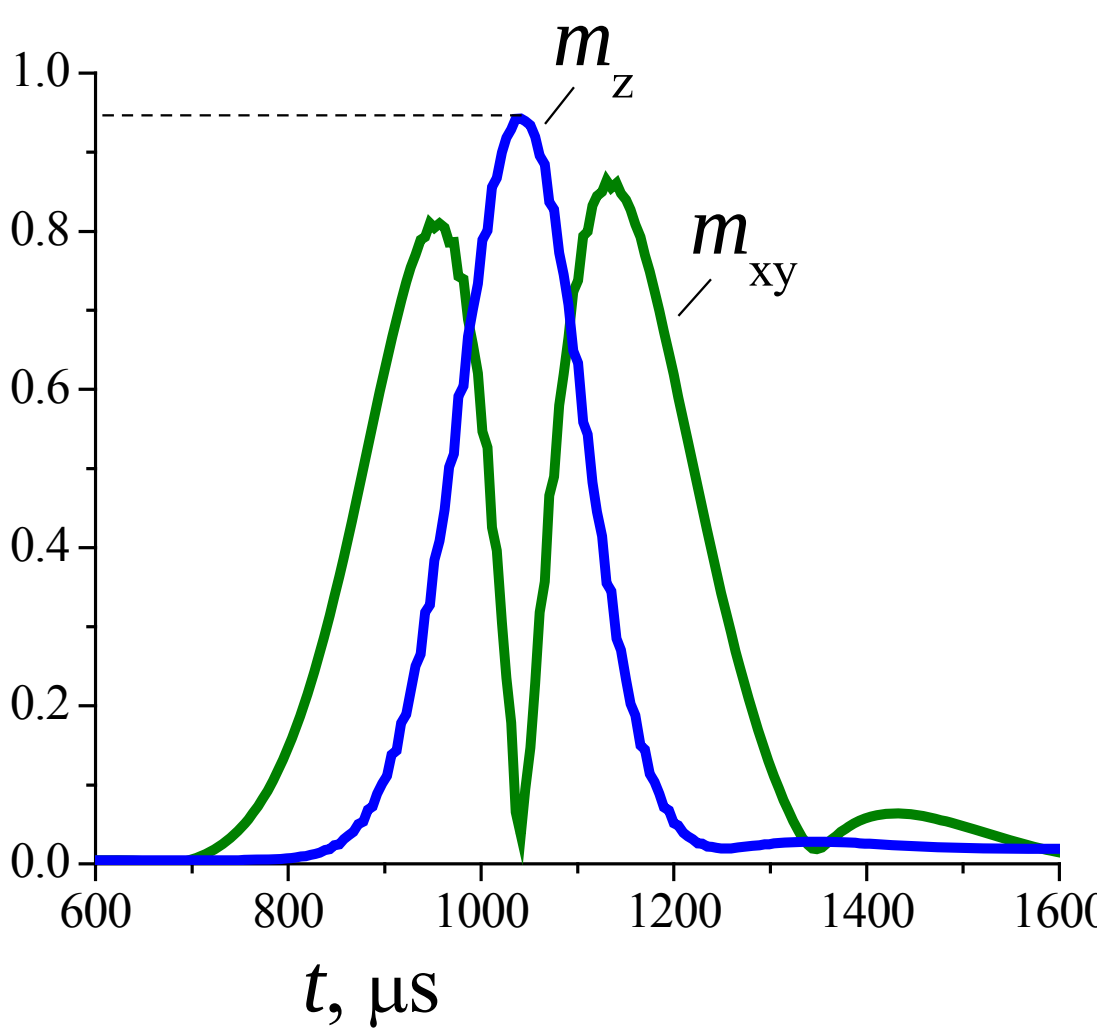

Fig. 9. Dynamics of the system for $\tau_{echo} \approx 0.$ and $L \approx 0.008$. The value of $m_{z,\max}$ is $m_{z,\max}$

Now we verify the correctness of the relation (10) which states that $L_{opt}$ does not depend on the resistance of the circuit, $R$. For this, we changed the quality factor, $Q$, by a factor в 3 (decreased the resistance, $R$, by a factor three) at $L = 0.075H$. (The quality factor of the LCRS circuit is given by $Q = \langle b_0 \rangle L / R$ r as $\langle b_0 \rangle$ is equal to the resonant frequency of the circuit.) The dynamics of the system, which is shown in Fig. 7, remained practically unchanged in the most important time-interval. If, in addition to the above described modification, we decrease by a factor of three parameter, $\alpha$, and simultaneously decrease by a factor of three the inductance, then we obtain the same results (the dependences shown by dotted lines in Fig. 7).

Finally, we will check relation (10) by changing the duration of the spin echo. For this we increase the parameter of the Gaussian distribution from $\Gamma = 3.5 \times 10^3 s^{-1}$ to $\Gamma = 10^4 s^{-1}$. The spin echo signal of the reduced duration is shown in Fig. 8.

For $\tau_{echo} \approx 0.325ms$ from formula (10) we obtain $L_{opt} \approx 0.008$. Fig. 9 demonstrates the dynamics of the system for $\tau_{echo} \approx 0.325ms$ and $L \approx 0.008$. One can see again that formula (10) provides a reliable estimate for the optimal inductance.

Certainly, the current in the coil 2 should be terminated when the longitudinal nuclear magnetization reaches its maximal value. As an example, in Fig. 4 the current should be terminated at $1160\mu s$ after the end of the second $\pi$-pulse.

## 4. Conclusion

We considered here the "dynamical" regime of the fast radiation damping (FRD). We have shown that FRD during the process of the spin echo formation can effectively restore the longitudinal nuclear magnetization. We have obtained an estimate for the optimal choice of the circuit parameters for the FRD, and confirmed it in our numerical experiments.

### Acknowledgement

This work was carried out under the auspices of the NNSA of the U. S. DOE at LANL under Contract No. DEAC52-06NA25396. We thank the LDRD program at LANL for funding this research.

### Appendix

We will consider the time interval from the beginning of the echo formation, $t'$, until the instant, $t''$, when the longitudinal magnetization reaches its maximum value. (See the left green peak in Fig. 6.) For our set of parameters, the time constant of the LCRS circuit (relaxation time) $2L/R$ is much greater than the time of the echo formation, $\tau_{echo}$, so we can safely put $R=0$ in Eqs (6). We will express $\langle m_x \rangle$ and the current, $I$, in terms of their amplitudes, $f = f(t)$, and $J = J(t)$:

$$\begin{aligned} \langle m_x \rangle &= f \exp\left(i\langle b_0 \rangle t\right), \\ I &= J \exp\left(i\langle b_0 \rangle t\right). \end{aligned} \tag{A1}$$

Substituting these expressions into the last equation in (6), we obtain the equation for the amplitudes:

$$\ddot{J} + 2i\langle b_0 \rangle \dot{J} = -(\alpha / L)\left( \ddot{f} + 2i\langle b_0 \rangle \dot{f} - \langle b_0 \rangle^2 f \right). \tag{A2}$$

Taking into consideration that the characteristic time of the amplitude variation is much longer than the period of the Larmor precession, $2\pi / \langle b_0 \rangle$, we can reduce this equation to

$$2i\langle b_0 \rangle \dot{J} = (\alpha / L)\langle b_0 \rangle^2 f . \tag{A3}$$

This equation gives the relation between the amplitude of the transverse magnetization and the derivative of the amplitude of current,

$$\dot{J} = -i\left(\alpha\left\langle b_0\right\rangle / 2L\right) f \,. \tag{A4}$$

Integrating (A4) over the time interval (from $t'$ to $t''$) we obtain,

$$J(t'') = -i\left(\alpha\left\langle b_0\right\rangle / 2L\right)\int_{t'}^{t''} f(t)dt \,. \tag{A5}$$

The integral in (A5) is, obviously, the area below the green peak in Fig. 6. We approximate this area with a triangle of height 1, midline $\tau$, and base $t''-t'=2\tau$. Then, the integral in Eq. (5) equals $\tau$. It follows from Eq. (A5) and Eqs (6) that during the considered time-interval the nuclear magnetization experiences a linearly polarized *rf* pulse of maximum amplitude,

$$\beta\alpha\left\langle b_0\right\rangle \tau / 2L. \tag{A6}$$

Approximating this pulse as a triangle (see Fig. 6) with the base, $2\tau$, and height (A6), we estimate its dimensionless area as $\beta\alpha\left\langle b_0\right\rangle \tau^2 / 2L$. The area of the corresponding pulse of the circular polarized field is $\beta\alpha\left\langle b_0\right\rangle \tau^2 / 4L$. Next, we approximate $\tau^2$ as a half of the corresponding value $\tau^2_{echo}$ of the unperturbed echo (with no radiation damping): $\tau^2 = \tau^2_{echo}/2$ (see Fig. 6). For optimal radiation damping, the area of the circular polarized pulse, which is equal to the angle of rotation of the nuclear magnetization, should be equal to $\pi/2$. From equation,

$$\beta\alpha\left\langle b_0\right\rangle \tau^2_{echo} / 8L = \pi / 2 \,, \tag{A7}$$

we obtain condition (10).